\documentclass[runningheads]{llncs}
\usepackage[T1]{fontenc}
\usepackage{graphicx,verbatim}
\usepackage{hyperref}
\usepackage{amsmath}
\usepackage{amssymb}
\usepackage{booktabs}
\usepackage{multirow}
\usepackage{rotating} 
\usepackage{subcaption}
\usepackage{pgffor}
\usepackage{orcidlink}
\begin{document}
\title{Diffusing the Blind Spot: Uterine MRI Synthesis with Diffusion Models}

\author{Johanna P. M\"uller\inst{1}\orcidlink{0000-0001-8636-7986} \and
Anika Knupfer \inst{1} \and
Pedro Blöss \inst{1} \and
Edoardo Berardi Vittur \inst{1} \and
Bernhard Kainz\inst{1,2}\orcidlink{0000-0002-7813-5023} \and
Jana Hutter\inst{1}\orcidlink{0000-0003-3476-3500}
}

\authorrunning{M\"uller et al.}
\institute{Friedrich–Alexander University Erlangen–N\"urnberg, DE \\
\email{johanna.paula.mueller@fau.de} \\
 \and Imperial College London, London, UK 
}

\maketitle              % typeset the header of the contribution
\begin{abstract}
Despite significant progress in generative modelling, existing diffusion models often struggle to produce anatomically precise female pelvic images, limiting their application in gynaecological imaging, where data scarcity and patient privacy concerns are critical. To overcome these barriers, we introduce a novel diffusion-based framework for uterine MRI synthesis, integrating both unconditional and conditioned Denoising Diffusion Probabilistic Models (DDPMs) and Latent Diffusion Models (LDMs) in 2D and 3D. Our approach generates anatomically coherent, high fidelity synthetic images that closely mimic real scans and provide valuable resources for training robust diagnostic models. We evaluate generative quality using advanced perceptual and distributional metrics, benchmarking against standard reconstruction methods, and demonstrate substantial gains in diagnostic accuracy on a key classification task. A blinded expert evaluation further validates the clinical realism of our synthetic images. We release our models with privacy safeguards and a comprehensive synthetic uterine MRI dataset to support reproducible research and advance equitable AI in gynaecology. The code and data are available at \url{https://github.com/ividja/SynthUterus}.
% Authors must provide keywords and are not allowed to remove this Keyword section.

\end{abstract}
\section{Introduction}
Generative models, particularly diffusion-based architectures, have demonstrated remarkable success across a wide range of applications in computer vision and medical imaging. However, despite their potential, key anatomical structures, such as the uterus and female pelvis, remain conspicuously absent from most publicly available models. While gynaecologists rely on their clinical expertise for diagnosis and treatment, making the interpretation process highly observer-dependent, the lack of high-quality uterine MRI datasets limits the development of tools that can support clinical education and improve diagnostic accuracy. Generative methods can be essential not only for training and reducing bias but also for enhancing the ability to detect complex or rare conditions like fibroids, adenomyosis, and congenital uterine anomalies. The scarcity of comprehensive uterine imaging datasets, compounded by privacy concerns, has hindered the development of robust diagnostic tools for these critical conditions. Given the high variability of female pelvic anatomy between individuals, by providing more diverse and representative data, generative models can help create diagnostic tools that assist clinicians in making faster, more accurate decisions, ultimately leading to improved patient outcomes. 

From a machine learning perspective, this represents an opportunity to leverage the power of deep generative models, such as Denoising Diffusion Probabilistic Models (DDPMs) and Latent Diffusion Models (LDMs), to fill this gap. Diffusion models are particularly well-suited to medical image synthesis due to their ability to generate high-quality, anatomically realistic images by learning complex distributions from limited data. For gynaecologists, access to synthetic yet anatomically realistic uterine MRI scans can aid diagnosis by facilitating anomaly comparison and strengthening AI models trained on limited data, thereby supporting clinical workflows in data-scarce settings.

\noindent\textbf{Contributions.} In this work, we present a novel generative framework for uterine MRI synthesis, addressing the need for both data augmentation and the generation of anatomically correct images for clinical use. 
Our contributions include: (1) the development of a tailored approach for synthesising uterine MRIs with diffusion models, in 2D and 3D, (2) the introduction of both unconditional and conditioned models that enable generation of diverse uterine anatomies, (3) evaluation on a clinically relevant task such as classification.

\section{Related Work}

\noindent\textbf{Uterus Imaging Datasets.}
Imaging plays a vital role in gynaecology and medical AI, yet publicly available datasets focused on the female pelvis, particularly high-resolution MRI, remain limited. Datasets such as UterUS~\cite{bones2024automatic} concentrate on transvaginal ultrasound and lack MRI data from adult, non-pregnant patients, omitting the pathological diversity needed for clinical relevance. The UMD dataset~\cite{Li2024} represents a major advance, providing annotated sagittal T2-weighted pelvic MRIs with histologically confirmed uterine myomas, segmentations, and FIGO classifications to support diagnosis and treatment planning. However, it largely comprises pathological cases, limiting the utility of models that depend on normal anatomy for weakly-, self-, or unsupervised learning. Without sufficient healthy examples, such methods struggle to differentiate typical from atypical presentations, reducing clinical reliability and generalisability. Additional datasets like the \href{https://zenodo.org/records/10979813}{Intrapartum Ultrasound Grand Challenge 2024} and the \href{https://www.cancerimagingarchive.net/collection/tcga-ucec/}{TCGA Uterine Corpus Endometrial Carcinoma Collection} are highly specialised, highlighting the ongoing lack of comprehensive, balanced datasets covering both healthy and pathological uterine anatomy across imaging modalities.

\noindent\textbf{Diffusion Models in Medical Imaging.}
Diffusion models have recently emerg-ed as powerful generative tools in medical imaging, enabling stable training and high-quality, anatomically coherent image synthesis. They have been applied successfully in brain MRI~\cite{pinaya2022brain,dorjsembe2024conditional}, chest CT~\cite{liu2023utilizing}, and digital pathology~\cite{pozzi2024generating} for image generation, inpainting, and data augmentation. These methods enable the creation of realistic synthetic datasets that support downstream tasks such as classification, reconstruction and anomaly detection~\cite{kazerouni2023diffusion,yang2023diffmic,webber2024diffusion,behrendt2024leveraging,baugh2024image}. However, their use in pelvic and gynaecological MRI remains limited due to scarce publicly available datasets of uterine anatomy, with particularly few examples of healthy patients. Expanding diffusion-based synthetic data generation in this area could address data scarcity, reduce annotation demands, and facilitate robust AI development.

\section{Method}

\noindent\textbf{Denoising Diffusion Probabilistic Models (DDPMs).}  
We model the true data distribution $p_{\text{data}}(x)$ using a DDPM~\cite{ho2020denoising}, which learns to reverse a fixed noising process defined by:

\begin{equation}
q(x_t \mid x_0) = \mathcal{N}(x_t; \sqrt{\bar{\alpha}_t} x_0, (1 - \bar{\alpha}_t) \mathbf{I}),
\end{equation}

where $x_t$ is a noisy version of the input image $x_0$ at diffusion step $t$, and $\bar{\alpha}_t$ is the cumulative product of variance schedule coefficients $\alpha_t$. The denoising model $p_\theta(x_{t-1} \mid x_t)$ is parameterised by a U-Net with time-step embeddings and spatial self-attention. We minimise the DDPM loss:

\begin{equation}
\mathcal{L}_{\text{DDPM}}(\theta) = \mathbb{E}_{x_0, \epsilon, t} \left[ \| \epsilon - \epsilon_\theta(x_t, t, c) \|_2^2 \right],
\end{equation}

where \(x_0\) is the original clean image, \(\epsilon_0 \sim \mathcal{N}(0, \mathbf{I})\) is the initial Gaussian noise added to \(x_0\), \(\epsilon\) is the noise added at timestep \(t\), \(x_t\) is the noisy image at timestep \(t\), \(c\) denotes conditioning information such as class labels or segmentation maps, and \(\epsilon_\theta(x_t, t, c)\) is the model's predicted noise.

\noindent\textbf{Latent Diffusion Models (LDMs).}  
To scale the generative process to high-resolution outputs efficiently, we incorporated Latent Diffusion Models (LDMs) ~\cite{rombach2022high} for final-stage refinement. LDMs operate in a learned latent space $\mathcal{Z} \subset \mathbb{R}^{h \times w \times c}$ rather than the pixel space $\mathcal{X}$. A convolutional autoencoder $(\mathcal{E}, \mathcal{D})$ was trained to minimise:

\begin{equation}
\mathcal{L}_{\text{VAE}}(\phi, \psi) = \mathbb{E}_{x \sim p_{\text{data}}} \left[ \| x - \mathcal{D}_\psi(\mathcal{E}_\phi(x)) \|_2^2 \right],
\end{equation}

ensuring that $\mathcal{E}_\phi(x) = z$ retains all clinically relevant uterine features.

The diffusion model then operates in latent space as:

\begin{equation}
z_t = \sqrt{\bar{\alpha}_t} z_0 + \sqrt{1 - \bar{\alpha}_t} \epsilon, \quad \epsilon \sim \mathcal{N}(0, \mathbf{I}),
\end{equation}

with loss function:

\begin{equation}
\mathcal{L}_{\text{LDM}}(\theta) = \mathbb{E}_{z_0, \epsilon, t} \left[ \| \epsilon - \epsilon_\theta(z_t, t, c) \|_2^2 \right],
\end{equation}

where $z_0 = \mathcal{E}(x)$ and $c$ again denotes conditioning inputs. Final reconstructions are obtained via $\hat{x} = \mathcal{D}(z_0)$.

\noindent\textbf{Preprocessing} %and Multi-resolution Pipeline.}
T2-weighted sagittal pelvic MRI scans were preprocessed to ensure anatomical consistency and facilitate multi-resolution modelling. Each volume $x \in \mathbb{R}^{H \times W \times S}$ was corrected for bias field inhomogeneity and standardised to zero mean and unit variance per scan. Using weakly supervised uterus localisation performed by a trained U-Net on a small set of annotated images, we extracted a region of interest (ROI) encompassing the uterus and adjacent structures. This ROI was then resampled to a standard in-plane resolution of $1.0$ mm.

\noindent\textbf{Text Conditioning.}
To enhance control and enable anatomically and clinically relevant synthesis, we incorporated text- and class-based conditioning into our diffusion models. Text conditioning uses structured natural language prompts, \emph{e.g.}, keywords for uterine position (anteflexed, retroflexed, anteverted, retroverted), MRI parameters (\emph{e.g.}, 1.5T, 3T), and sequence types (\emph{e.g.}, TSE, HASTE). The input $c$ is encoded via a pretrained text encoder (e.g., Transformer or CLIP), producing an embedding that modulates the denoising network via cross-attention. Class conditioning specifies categorical labels such as uterine position. This hybrid framework enables generation of anatomically plausible pelvic MRI slices and volumes aligned with clinical descriptors, supporting explicit control over synthesised image characteristics.

\noindent\textbf{Privacy Filtering.}
To mitigate risks of patient reidentification, especially for the full pelvic scans, and prevent overfitting through memorisation we implemented a post-hoc privacy filter for all generated images $\hat{x}$. Each $\hat{x}$ was embedded into a perceptual space using a frozen encoder $f: \mathcal{X} \rightarrow \mathbb{R}^d$, trained independently from the diffusion model. For each training image $x_i$, we computed the cosine similarity:

\begin{equation}
\text{sim}(\hat{x}, x_i) = \frac{f(\hat{x}) \cdot f(x_i)}{\|f(\hat{x})\| \, \|f(x_i)\|}.
\end{equation}

Generated samples were flagged and discarded if they exceeded a similarity threshold $\tau$ against any training image:

\begin{equation}
\max_i \, \text{sim}(\hat{x}, x_i) > \tau, \quad \text{with} \quad \tau = 0.95.
\end{equation}
To detect higher-level near-duplicates, we compared structural embeddings from intermediate encoder layers and clustered them using approximate nearest neighbour search. This multi-scale filtering ensures accepted samples are sufficiently distinct from the training data, supporting patient anonymity and adherence to generative privacy standards.

\section{Evaluation}
\begin{figure}[htbp]
  \centering
  \includegraphics[width=\textwidth]{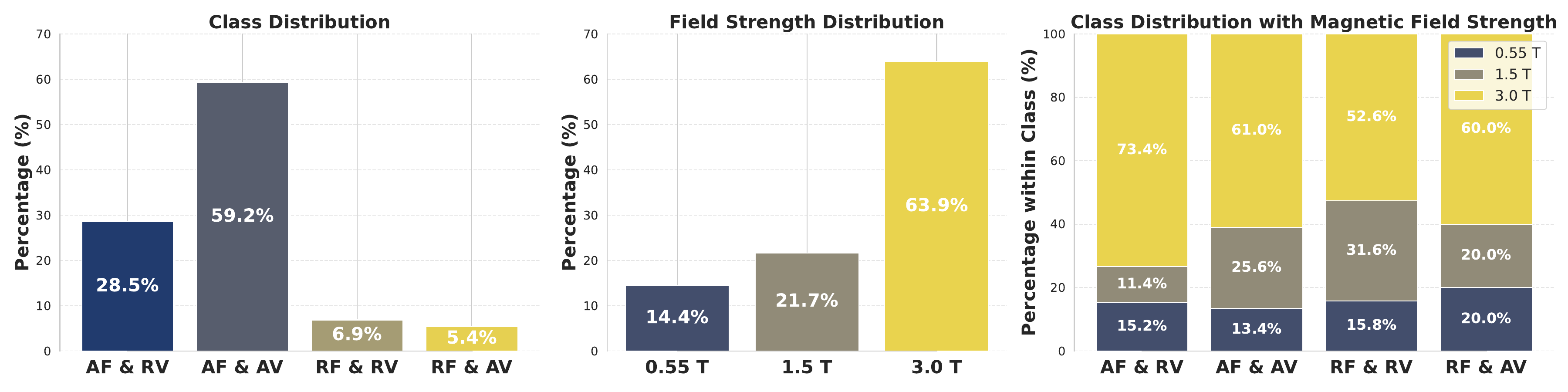}
  \caption{FUNDUS dataset composition. (l.) Distribution of anatomical classes based on uterine orientation combinations - Anteflexed (AF), Retroflexed (RF), Anteverted (AV), and Retroverted (RV). (m.) Distribution of scanner magnetic field strengths (in Tesla). (r.) Breakdown of each anatomical class by scanner field strength.}
  \label{fig:fundus_composition_full}
\end{figure}

\noindent\textbf{Datasets.}
The UMD~\cite{pan2024large} dataset consists of sagittal T2-weighted pelvic MRI scans from 300 patients (ages $21-86$) with histologically confirmed uterine myomas, acquired on a Philips 3T system. Pixel-level annotations were provided by experienced gynaecologists and radiologists for the uterine cavity, wall, myomas, and nabothian cysts. Each case is labelled according to FIGO classification (types 0–8). Images and masks are provided in NIfTI format and are publicly available via \href{https://figshare.com/articles/dataset/UMD_zip/23541312?file=44111183}{Figshare}.
The in-house FUNDUS dataset consists of 267 T2-weighted sagittal pelvic MRI scans of healthy individuals collected retrospectively at the University Hospital Erlangen (UKER), Germany. The age of the patients ranges from 11 to 82 years. The dataset is characterised by its variability in imaging parameters due to the lack of a standardised protocol for MRI of the abdomen and pelvis. These include differences in field strength (0.55 T, 1.5 T, 3 T), scanner type (Siemens, PHILIPS), resolution ($208 - 832$), sequence (TSE, HASTE) and use of contrast agents. In addition, natural anatomical variations during the menstrual cycle were recorded. Some individuals were scanned multiple times, revealing changes due to menstrual phase, bladder filling or age, further increasing the diversity of the dataset.

\noindent\textbf{Metrics.}
Reconstruction quality (for encoders) and generation quality (for diffusion models) were evaluated using Learned Perceptual Image Patch Similarity (LPIPS) and Fréchet Inception Distance (FID). LPIPS quantifies perceptual similarity between individual image patches, capturing subtle, fine-grained differences, while FID compares the overall distributions of real and synthetic images to assess dataset-level realism. For both metrics, lower values indicate higher quality. Classification performance was measured using the Area Under the Receiver Operating Characteristic Curve (AUC) and macro-averaged F1-score (F1), reflecting discriminative ability and balanced class performance.

\noindent\textbf{Training and Hyperparameters.}
All models were trained on NVIDIA $A100$ GPUs ($40 - 80$ GB memory). DDPMs followed the implementation from~\cite{von-platen-etal-2022-diffusers}, and LDMs used the framework by~\cite{reynaud2024echonet} with a Variational Autoencoder (VAE) with a 16× compression ratio and an EDM U-Net backbone~\cite{Karras2022edm}. Models were trained for up to $2000$ epochs with early stopping based on validation loss (patience: $50$) and class-weighted sampling. We used the AdamW optimiser with learning rates in $[1\mathrm{e}{-5}, 1\mathrm{e}{-3}]$ and batch sizes between $1$ and $64$ (126 for Latent U-Net), depending on model size and GPU memory. Diffusion models used 1000 denoising steps with discrete schedules. Both DDPMs and LDMs used a perceptual loss weighting $\lambda_{\text{LPIPS}} \in [0.1, 1.0]$. Text and class conditioning used dropout rates sampled from $[0.0, 0.2]$. All hyperparameters were tuned via grid search on a held-out validation split.

\begin{figure}[htbp]
  \centering
  \setlength{\tabcolsep}{0pt} % remove horizontal padding between columns
  \renewcommand{\arraystretch}{1} % vertical spacing in cells

  \begin{minipage}[t]{0.51\textwidth}
    \centering
    \resizebox{\textwidth}{!}{
    \begin{tabular}{p{0.3cm}cccc}
      & \multicolumn{4}{c}{\textbf{2D}} \\
      & AF \& AV & RF \& AV & AF \& RV & RF \& RV \\

      \rotatebox{90}{\small Class} &
      \includegraphics[width=.245\textwidth]{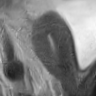} &
      \includegraphics[width=.245\textwidth]{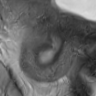} &
      \includegraphics[width=.245\textwidth]{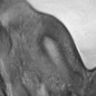} &
      \includegraphics[width=.245\textwidth]{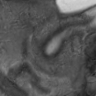} \\

      \rotatebox{90}{\small + Text} &
      \includegraphics[width=.245\textwidth]{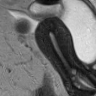} &
      \includegraphics[width=.245\textwidth]{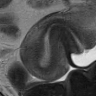} &
      \includegraphics[width=.245\textwidth]{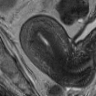} &
      \includegraphics[width=.245\textwidth]{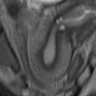} \\
    \end{tabular}
    }
  \end{minipage}%
  \hspace{0pt}% no space between minipages
  \begin{minipage}[t]{0.49\textwidth}
    \centering
    \resizebox{\textwidth}{!}{
    \begin{tabular}{ccccc}
      & \multicolumn{4}{c}{\textbf{3D}} \\
      & AF \& AV & RF \& AV & AF \& RV & RF \& RV \\

      &
      \includegraphics[width=.245\textwidth]{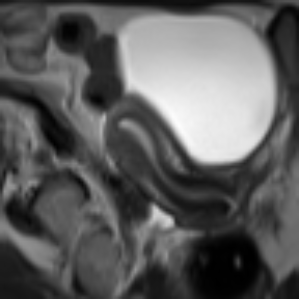} &
      \includegraphics[width=.245\textwidth]{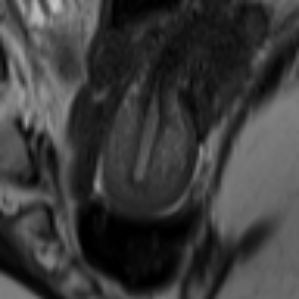} &
      \includegraphics[width=.245\textwidth]{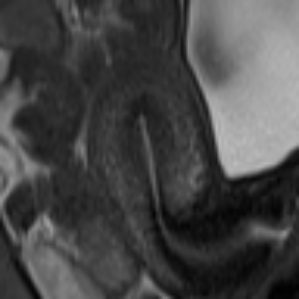} &
      \includegraphics[width=.245\textwidth]{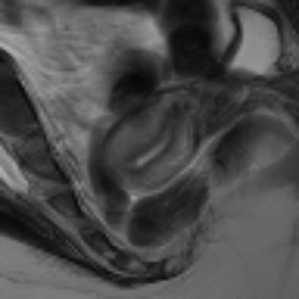} \\

      &
      \includegraphics[width=.245\textwidth]{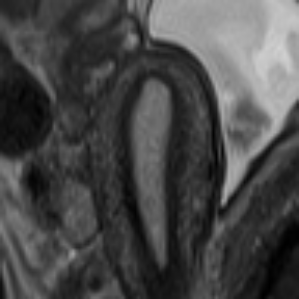} &
      \includegraphics[width=.245\textwidth]{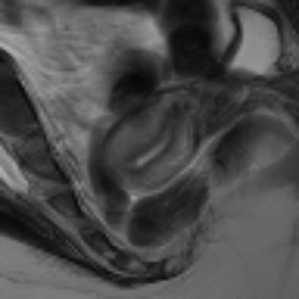} &
      \includegraphics[width=.245\textwidth]{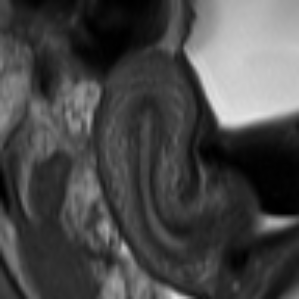} &
      \includegraphics[width=.245\textwidth]{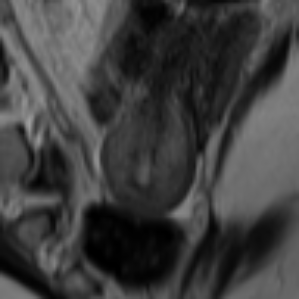} \\
    \end{tabular}
    }
  \end{minipage}
  
  \caption{Generated images from class-conditioned, and class- and text-conditioned DDPMs for 2D (left) and 3D (right) models. Uteri are shown in four orientation combinations: Anteflexed (AF), Retroflexed (RF), Anteverted (AV), and Retroverted (RV).}
  \label{fig:class_text_comparison}
\end{figure}

\noindent\textbf{Downstream Clinical Task.}  
We evaluated classification using a 2D ResNet-18 under multiple regimes: fully supervised on the ground truth (GT) dataset FUNDUS, supervised with a pretrained ResNet-18, weakly supervised with only $10~\%$ labelled data, and unsupervised via k-means clustering. These regimes were also applied to our synthetic datasets SynthUterus and SynthUterus (ROI), generated by class- and text-conditioned DDPMs capturing uterine positions and magnetic field strength of the scanners. Models were optimised with cross-entropy loss, producing softmax-normalised outputs, and evaluated on a held-out test set.

\subsection{Results and Discussion}

\begin{figure}[!ht]
    \centering
    \includegraphics[width=\textwidth]{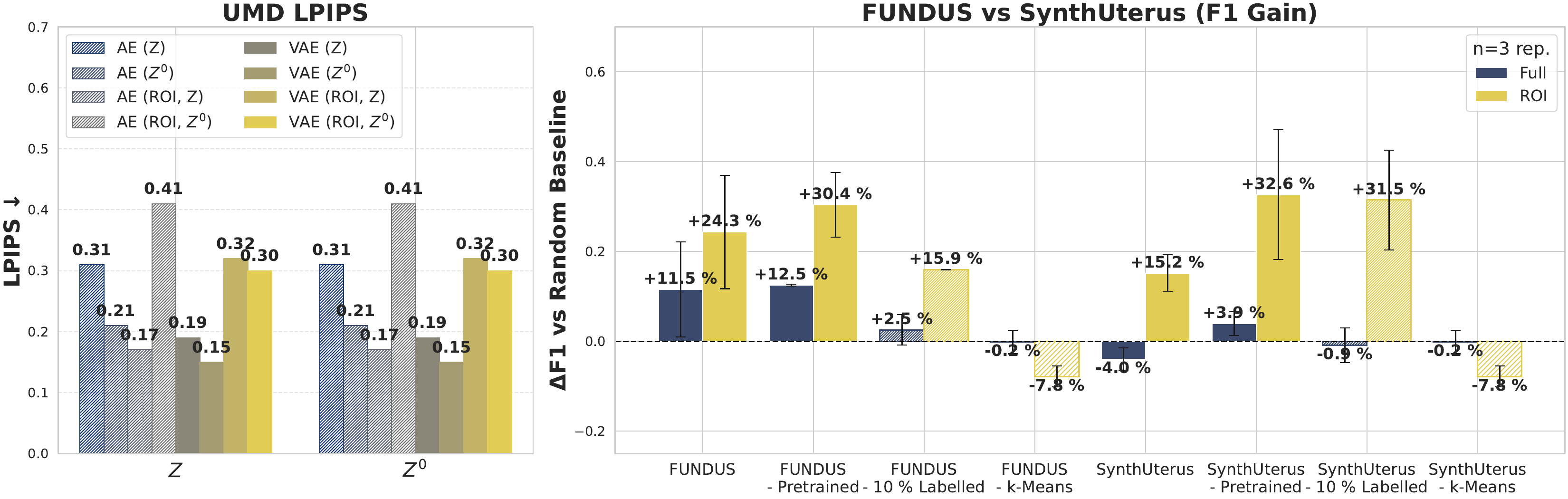}
    \caption{
        Perceptual reconstruction quality (LPIPS) and classification (position) performance gain ($\Delta$ F1) across preprocessing strategies. Left: LPIPS scores for AE and VAE encoder models on the UMD test set under varying training preprocessing setups, tested on all slices of the volume and only the central slices. Lower values indicate better perceptual similarity. Right: $\Delta$ F1 relative to a random baseline on the FUNDUS and SynthUterus datasets using different training strategies. Preprocessing abbreviations: Z – full volume; Z$^0$ – central 3 slices; ROI – cropping to the uterus. 
    }
    \label{fig:image-reconstruction-metrics}
\end{figure}

\noindent\textbf{Image Reconstruction and Generation.} 
Fig.~\ref{fig:image-reconstruction-metrics} (left): Using LPIPS (AlexNet), the AE trained on FUNDUS achieved a score of $0.17$ on both full volumes and central slices (Z$^0$), while the VAE reached $0.15$. Applying ROI cropping to FUNDUS increased LPIPS to $0.41$ for the AE and $0.30$ for the VAE. All UMD inputs were evaluated without cropping to ROI.
We evaluated 2D and 3D DDPMs using FID and LPIPS across uterine orientation classes and conditioning setups (Tab.~\ref{tab:image-generation-per-class}): class only, class + ROI, class + text (C+T), and C+T + ROI.  Example images for qualitative evaluation are shown in Fig.~\ref{fig:class_text_comparison}. All 2D models were trained on the central slices for evaluation, trained on all slices in the volume, FID and LPIPS increased by $10~\%$ at minimum. Text-conditioned models without class-conditioning performed worse than class-only conditioned models in an extended ablation study. The 2D DDPM with C+T + ROI conditioning consistently achieved the best results. ROI cropping alone also improved performance, especially when combined with semantic input. In 3D, the best results came from class + ROI, though overall quality lagged behind 2D models. In Tab.~\ref{tab:image-generation-avg}, the ablation study shows that conditioning with class and text information combined with ROI preprocessing consistently improves image quality across DDPM and LDM models, with 2D LDMs achieving the best overall FID and LP scores.

\begin{table}[!ht]
\centering
\caption{
Ablation study on image generation quality across DDPM models and conditioning strategies by uterine orientation. 
FID: Fréchet Inception Distance; 
LP: Learned Perceptual Image Patch Similarity. Preprocessing as above.
\textbf{1st-ranked}, \underline{2nd-ranked} model configuration, individually for 2D and 3D.
}
\label{tab:image-generation-per-class}
\begin{tabular}{c l c c 
                *{8}{c}} 
& & & 
& \multicolumn{2}{c}{\textbf{AF \& AV}} 
& \multicolumn{2}{c}{\textbf{RF \& AV}} 
& \multicolumn{2}{c}{\textbf{AF \& RV}} 
& \multicolumn{2}{c}{\textbf{RF \& RV}} \\
\cmidrule(lr){5-6} \cmidrule(lr){7-8} \cmidrule(lr){9-10} \cmidrule(lr){11-12}
& \textbf{Model} & \textbf{ROI} & \textbf{Z$^0$} 
& FID$\downarrow$ & LP$\downarrow$ 
& FID$\downarrow$ & LP$\downarrow$ 
& FID$\downarrow$ & LP$\downarrow$ 
& FID$\downarrow$ & LP$\downarrow$ \\
\cmidrule(lr){5-5} \cmidrule(lr){6-6}
\cmidrule(lr){7-7} \cmidrule(lr){8-8}
\cmidrule(lr){9-9} \cmidrule(lr){10-10}
\cmidrule(lr){11-11} \cmidrule(lr){12-12}
% --- DDPM (2D) ---
\multirow{4}{*}{\rotatebox{90}{2D}} 
& + Class         &     -        & \checkmark  & $7.89$ & $0.52$ & $7.46$ & $0.52$ & $7.55$ & $0.50$ & $8.16$ & $0.51$ \\
&           & \checkmark  & \checkmark  & $\underline{3.42}$ & $\mathbf{0.38}$ & $\underline{2.80}$ & $\underline{0.38}$ & $\underline{2.61}$ & $\underline{0.38}$ & $\underline{2.19}$ & $\underline{0.40}$ \\
&  + C+T          &        -     & \checkmark  & $4.09$ & $0.48$ & $3.18$ & $0.47$ & $4.00$ & $0.48$ & $4.47$ & $0.48$ \\
&           & \checkmark  & \checkmark  & $\mathbf{1.05}$ & $\underline{0.40}$ & $\mathbf{0.33}$ & $\mathbf{0.37}$ & $\mathbf{0.25}$ & $\mathbf{0.37}$ & $\mathbf{0.65}$ & $\mathbf{0.38}$ \\
% --- DDPM (3D) ---
\midrule
\multirow{4}{*}{\rotatebox{90}{3D}}
& + Class   &       -      &  - & $27.12$ & $0.72$ & $\underline{24.88}$ & $\underline{0.71}$ & $25.77$ & $0.71$ & $\mathbf{24.09}$ & $0.71$ \\
&           & \checkmark  & -  & $\underline{24.66}$ & $\mathbf{0.68}$ & $\mathbf{24.13}$ & $\mathbf{0.70}$ & $\mathbf{23.61}$ & $\mathbf{0.69}$ & $24.60$ & $\underline{0.70}$ \\
& + C+T     &  -           & -  & $26.11$ & $0.72$ & $27.46$ & $0.72$ & $\underline{24.33}$ & $0.71$ & $24.78$ & $0.71$ \\
&           & \checkmark  & -  & $\mathbf{24.51}$ & $\underline{0.69}$ & $25.28$ & $\mathbf{0.70}$ & $24.77$ & $\underline{0.70}$ & $\underline{24.55}$ & $\mathbf{0.69}$ \\
\end{tabular}
\end{table}

\begin{table}[!ht]
\centering
\caption{
Ablation study and average evaluation scores of DDPMs and LDMs across all uterine positions.  
FID: Fréchet Inception Distance;  
LP: Learned Perceptual Image Patch Similarity. Preprocessing as above.
\textbf{1st-ranked}, \underline{2nd-ranked} configuration for each model.
}
\label{tab:image-generation-avg}
\begin{tabular}{l c c 
                c c 
                c c 
                c c}
% Model group headers

&&
& \multicolumn{2}{c}{\textbf{DDPM (2D)}} 
& \multicolumn{2}{c}{\textbf{LDM (2D)}} 
& \multicolumn{2}{c}{\textbf{DDPM (3D)}} \\
\cmidrule(lr){4-5} \cmidrule(lr){6-7} \cmidrule(lr){8-9}
\textbf{Cond.} 
& \textbf{ROI} & \textbf{Z$^0$} 
& FID$\downarrow$ & LP$\downarrow$ 
& FID$\downarrow$ & LP$\downarrow$ 
& FID$\downarrow$ & LP$\downarrow$ \\
% Individual metric column rules
\cmidrule(lr){4-4} \cmidrule(lr){5-5}
\cmidrule(lr){6-6} \cmidrule(lr){7-7}
\cmidrule(lr){8-8} \cmidrule(lr){9-9}
Uncond.         & -         & \checkmark & $8.46$ & $0.45$ & $3.44$ & $0.42$ & $27.03$ & $0.72$ \\
                & \checkmark   & \checkmark & $\underline{1.90}$ & $\mathbf{0.37}$ & $2.17$ & $\underline{0.35}$ & $25.45$ & $0.70$ \\
+ Class         & -          & \checkmark & $7.77$ & $0.51$ & $2.13$ & $0.39$  & $25.46$ & $0.71$ \\
                & \checkmark   & \checkmark & $2.76$ & $0.39$ & $\underline{1.45}$ & $0.53$  & $\mathbf{24.25}$ & $\mathbf{0.69}$ \\
+ C+T           & -          & \checkmark & $3.93$ & $0.48$ & $1.97$ & $0.43$  & $25.67$ & $0.72$ \\
                & \checkmark   & \checkmark & $\mathbf{0.57}$ & $\underline{0.38}$ & $\mathbf{1.35}$ & $\mathbf{0.32}$  & $\underline{24.78}$ & $\underline{0.70}$ \\
\end{tabular}
\end{table}

\noindent\textbf{Synthetic Datasets.} 
The SynthUterus datasets include $800$ scans with $200$ synthetic images per class for each uterine position and are balanced to match the FUNDUS dataset distribution (Fig.~\ref{fig:fundus_composition_full}). Two versions were generated using class and text conditioned DDPMs: full images referred to as SynthUterus and uterus-focused region of interest crops referred to as SynthUterus ROI, capturing semantic and spatial details to improve training. Ten real and ten synthetic healthy pelvic ROI MRI samples were classified by three groups: non-expert AI researchers, less experienced radiologists and experienced pelvic radiologists. Their accuracies were $46.3\%$, $40\%$ and $50\%$ respectively, showing limited ability to distinguish real from generated images.

\noindent\textbf{Image Classification.} 
We evaluated classification performance across four training regimes: full supervision, pretrained ResNet-18, weak supervision with $10~\%$ labelled data, and unsupervised k-means, using both FUNDUS and SynthUterus datasets. Performance was reported in terms of improvement in F1 score over a random baseline on the FUNDUS test set, with $n=3$ repetitions, see Fig.~\ref{fig:image-reconstruction-metrics} (right). Models trained on SynthUterus ROI, consistently outperformed those trained on FUNDUS in weak-supervision settings, achieving a $+32.6~\%$ gain with $10~\%$ labelled data over Random, compared to $+15.9~\%$ for FUNDUS. Even under full supervision, SynthUterus achieved a modest boost ($+2.5~\%$) over FUNDUS if Resnet-18 was pretrained. The fully unsupervised k-Means clustering equally performed worse for both true and generated datasets.

\noindent\textbf{Discussion.}
Our results demonstrate that semantic and spatial conditioning significantly enhance 2D diffusion-based MRI synthesis, enabling the production of anatomically coherent and high-quality images. Notably, the synthetic ROI dataset improved classification robustness and, in some cases, surpass models trained on real data under weak supervision and supervised with pretrained encoders. This underlines the potential of diffusion-generated data to support clinically relevant tasks, particularly where annotated data is scarce.
While 3D DDPMs show promise, their performance is currently limited by longer training times and higher memory demands. Latent diffusion models remain sensitive to architectural choices; replacing the latent U-Net denoiser with transformer-based alternatives could improve anatomical fidelity and image realism.
Nonetheless, both expert assessments and downstream evaluations reveal the potential for shortcut learning, where models might rely on superficial or spurious image features instead of meaningful anatomical structures. This highlights the critical need for robust validation, especially on held-out and multicentre datasets, to ensure generalisability and clinical relevance. Additionally, employing a standard pretrained encoder such as SwAV~\cite{caron2020unsupervised} resulted in a mean Image Retrieval Score $IRS_{\infty,\alpha}$~\cite{Dombrowski_2025_CVPR} below $0.10$, indicating strong similarity among generated images. This may reflect a limited capacity of the encoder to distinguish between subtle uterine features and emphasise the need for higher diversity in image generation.

\section{Conclusion} 
We present a diffusion-based framework for synthetic pelvic MRI generation conditioned on uterine position and descriptive text, including scanner field strength. This approach enables scalable, privacy-preserving data augmentation to address limited annotations and patient confidentiality. Our conditioned 2D DDPM achieves state-of-the-art image quality, with synthetic data matching or surpassing real data performance in weakly and fully supervised settings, supporting robust model development in data-scarce scenarios. By releasing our pipeline and models, we aim to promote reproducible research and accelerate progress in this clinical domain.
Future work should focus on improving synthesis diversity through diversity modules and domain-specific encoders trained on multi-centre data, extending pathology conditioning to rare cases, incorporating radiology reports for richer conditioning, and rigorously evaluating privacy safeguards to ensure secure clinical deployment.

\begin{credits}
\emph{Acknowledgements:} The authors gratefully acknowledge the scientific support and HPC resources provided by the Erlangen National High Performance Computing Center (NHR@FAU) under the NHR projects b143dc and b180dc. NHR funding is provided by federal and Bavarian state authorities. NHR@FAU hardware is partially funded by the German Research Foundation (DFG) – 440719683. This work was supported by the ERC Project MIA-NORMAL 101083647 and the ERC Starting grant EARTHWORM 101165242, DFG 513220538 and 512819079, DFG Heisenberg 502024488 and by the state of Bavaria (HTA).
\end{credits}


\bibliographystyle{splncs04}
\bibliography{bibliography}

\end{document}